\documentclass[prl,preprint,showpacs,superscriptaddress,preprintnumbers]{revtex4}%
\usepackage{graphicx}
\usepackage{dcolumn}
\usepackage{bm}
\usepackage{amsmath}
\usepackage{amsfonts}
\usepackage{amssymb}

\newcommand\Figr[1]{Fig.~\ref{fig:#1}}
\newcommand\sinc{\operatorname{sinc}}
\newcommand*{\zoom}{1.6}

\begin{document}
\title{Two-dimensional droplet spreading over random topographical substrates}
\author{Nikos Savva}
\affiliation{Department of Chemical Engineering, Imperial College
London, London SW7 2AZ, UK}
\author{Serafim Kalliadasis}
\affiliation{Department of Chemical Engineering, Imperial College
London, London SW7 2AZ, UK}
\author{Grigorios A. Pavliotis}
\affiliation{Department of Mathematics, Imperial College London,
London SW7 2AZ, UK}
\date{\today }
\begin{abstract}
We examine theoretically the effects of random topographical
substrates on the motion of two-dimensional droplets via appropriate
statistical approaches. Different random substrate families are
represented as stationary random functions. The variance of the
droplet shift at both early times and in the long-time limit is
deduced and the droplet footprint is found to be a normal random
variable at all times. It is shown that substrate roughness
decreases droplet wetting, illustrating also the tendency of the
droplet to slide without spreading as equilibrium is approached. Our
theoretical predictions are verified by numerical experiments.
\end{abstract}
\maketitle

%================================================================================

Front propagation in heterogeneous media occurs in a wide variety of
areas in physics, ranging from transport phenomena in porous media
and reaction-diffusion-advection systems to crack propagation due to
lattice defects~\cite{Homsy87}. Heterogeneities are always present
in natural environments but quite often their precise form is
unknown. In such cases they can be modeled as random noise signals,
a more physical and practical assumption than e.g. periodic ones. An
example of front propagation central to interfacial hydrodynamics is
that of a moving contact line during liquid spreading on a solid
substrate, where heterogeneities usually originate from substrate
defects, either chemical~\cite{chemExp} or
topographical~\cite{roughExp}.

It is a fundamental problem to understand how random heterogeneities
influence the spreading dynamics and the characteristics of contact
line propagation, e.g. speed and location. Experimental studies on
droplet spreading -- a simple prototype for the study of contact
line motion -- suggest that substrates having highly irregular
micro-scale features, commonly called rough, can influence the
dynamics significantly~\cite{Cazabat86}. Several theoretical works
focused on equilibrium configurations and deterministic substrates:
the study by Wenzel~\cite{Wenzel36} on droplet equilibria on rough
substrates obtained an effective contact angle based on simplistic
energy arguments -- a formula which was supported by experiments in
a regime where the contact angles are not small~\cite{roughExp};
in~\cite{deGennes84} the contact line of a fluid wedge in the
presence of a single, localized defect was determined;
in~\cite{prevStudiesTheory} droplet equilibria on substrates with
regular, periodic features were considered; Cox~\cite{Cox83}
investigated multiple equilibria of an infinite fluid wedge on a
general non-periodic rough surface. Introducing randomness in the
variations of the substrates is clearly a realistic way to represent
roughness. There are a few studies in this direction but are based
on phenomenological modeling ideas as for example~\cite{Jansosns85}
which regarded rough substrate patches as contact angle point
sources or~\cite{Borgs95} which looked into the effects of surface
roughness using an Ising model. Studies on contact line dynamics on
random substrates were also based on phenomenological ideas,
e.g.~\cite{Mulinet} where postulated equations describing the
dynamics were utilized. Hence, to date a systematic fluid dynamics
treatment based on rational statistical approaches is still lacking.
As a result, contact line motion on heterogeneous substrates is far
from being well understood, unlike other problems in continuum
mechanics, such as porous media, which not only have been analyzed
from first principles but have also motivated new mathematical
approaches (cf. homogenization~\cite{Hornung}).

In this Letter we report the first detailed and systematic study of
the qualitative effects of random, small-scale spatial
heterogeneities on droplet motion, through the development of
appropriate statistical methodologies. The starting point is the
recent work in~\cite{SavvaKall09} on the motion of two-dimensional
(2D), partially wetting droplets over deterministic, spatially
heterogeneous substrates. The restriction to 2D implies that there
are no transverse variations and the contact line is essentially
treated as a set of two points. This simplified problem, albeit more
difficult to study experimentally, is a first step towards
understanding the influence of spatial heterogeneities, but the
required analysis still remains highly nontrivial. The model for the
droplet motion was obtained rigorously from the governing
hydrodynamic equations. Assuming small contact angles, a long-wave
expansion in the Stokes flow regime, yields a single equation for
the evolution of the droplet thickness $H\left( x,t\right)  $ over a
substrate $\eta\left( x\right)  $. In dimensionless form the
equation reads:
\begin{equation}
\partial_{t}H+\partial_{x}\left[  H^{2}\left(  H+\lambda\right)  \partial
_{x}^{3}\left(  H+\eta\right)  \right]  =0,\label{eq:Gov}%
\end{equation}
where $\lambda\ll 1$ is the non-dimensional slip length originating
from the Navier model imposed to alleviate the stress singularity
that occurs at the moving contact line~\cite{HuhScriven71}. Equation
\eqref{eq:Gov} describes the capillarity driven spreading that is
resisted by fluid viscosity: the term $\partial_t H$ results from
the viscous fluid motion and the term $\partial_x [\cdot]$
represents the effect of surface tension and also accounts for the
substrate curvature which contributes an additional capillary
pressure, $-\partial_x^2 \eta$. The spatial coordinate $x$, and time
$t$, are made non-dimensional by the characteristic length
$L=\sqrt{A/(2\alpha_s)}$ and time $\tau=3\mu
L/(\gamma\alpha_s^{3})$, respectively, where $A$ is the droplet
cross-sectional area, $\alpha_s$ is the equilibrium angle prescribed
by Young's law, $\mu$ is the viscosity and $\gamma$ is the surface
tension. $\eta(x)$ and $H(x)$ are scaled by $L\alpha_s$ and
$\lambda$ by $L\alpha_s/3$ respectively. Equation~(\ref{eq:Gov}) is
subject to a constant volume constraint and the boundary conditions
at the contact points that the droplet thickness vanishes and the
angle the free surface makes with the substrate remains equal to its
static value, $\alpha_s$. For quasi-static spreading, a singular
perturbation method was employed in~\cite{SavvaKall09} to
asymptotically match the solution in the bulk of the fluid with the
solution in the vicinity of the contact lines. This led to a set of
two integro-differential equations (IDEs) for the time-evolution of
the right and left contact points at $x=a_{\pm}\left(  t\right)$,
respectively. Their validity has been confirmed by comparisons with
numerical solutions to the full equation
in~(\ref{eq:Gov})~\cite{SavvaKall09}.

Here we use the set of IDEs to investigate the case where
$\eta\left( x\right) $ is a random function. We take
$|\eta\left( x\right)|  \ll1$ assuming also that its
variations occur at lengthscales that
are much longer than $\lambda$. The requirement of
substrate smoothness together with the fact that the `noise' is
spatial and enters the equations in a nonlinear fashion, precludes
casting the problem into the standard Langevin-Fokker-Planck
formalism, which is usually employed to study randomly perturbed
dynamical systems~\cite{Risken}.
\begin{figure}[ptb]
\includegraphics[scale=\zoom]{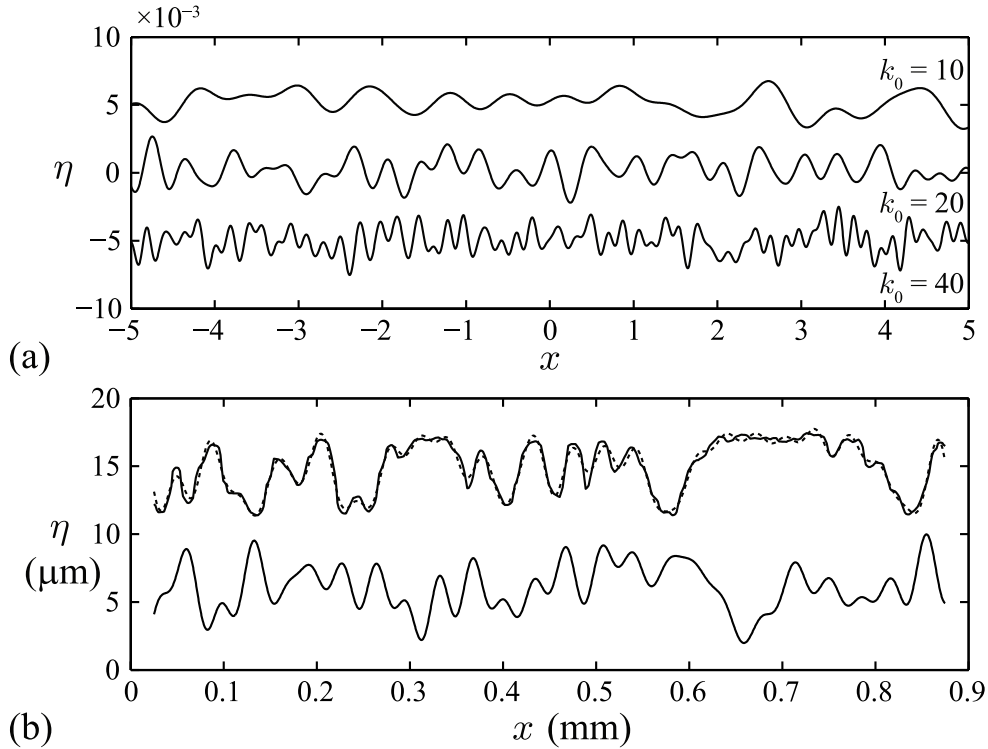}\caption{(a) Sample substrate realizations using
(\ref{eq:eta}) for $\eta_{0}=10^{-3}$ and $k_{0}=$ 10, 20 and 40.
(b) Top plot. Solid line: experimental profile from Hitchcock
\emph{et al.}~\cite{roughExp} for an alumina sample. Dashed line:
approximation obtained by projecting the experimental profile onto
$N=26$ harmonics in~\eqref{eq:eta}.
Lower plot: sample substrate realization for the same $\eta_0$ and $k_0$, but $N = 1000$.}%
\label{fig:01}%
\end{figure}

The primary fundamental difficulty with the substrate is the
development of a random representation that can have a large
frequency content and at the same time it is differentiable. A
convenient representation is the random function,
\begin{equation}
\eta\left(  x\right)  =\frac{\eta_{0}}{\sqrt{N}}\sum_{m=1}^{N}\alpha_{m}%
\sin\frac{k_{0}m}{N}x+\beta_{m}\cos\frac{k_{0}m}{N}x,\label{eq:eta}%
\end{equation}
where $\eta_{0}\ll1$ and $k_{0}$ are the characteristic amplitude
and wavenumber of the substrate, respectively, and $N$ is a large
positive integer. Here the coefficients $\alpha_{m}$ and $\beta_{m}$
are statistically independent normal variables with unit variance.
It is readily seen that $\eta\left(  x\right)  $ is a periodic
function with period $2\pi N/ k_{0}$, but we eventually take
$N\rightarrow\infty$ so that this periodicity is lost and $\eta(x)$
approaches a band-limited white noise (see \Figr{01}a; in this
limit, continuity of~(\ref{eq:eta}) and all its derivatives can be
carefully shown by Kolmogorov's continuity theorem). Representations
of this form, are also invoked to look into noise effects in other
contexts as they occur, for example, in electrical current signals
or in black-body radiation~\cite{Rice52}. 

An attractive feature of the stochastic representation
in~(\ref{eq:eta}) is that it generates an infinite family of
substrate realizations parameterized by two parameters, $\eta_{0}$
and $k_{0}$, which are often reported in experimental studies when
characterizing a rough substrate (in~\Figr{01}b, $\eta_0 \sim 0.78 -
2.1 \mu$m, wavelength $l_0 \sim 14 - 42 \mu$m). These two parameters may
be also be used to compute the `roughness coefficient of the substrate', $r$,
defined as the mean of the ratio of the actual surface area over its projected area. In
the limit of $\eta_{0}k_{0}\ll1$ we find $r=1+\eta_{0}^{2}k_{0}^{2}/6$, to be
contrasted with $r=1+\eta_{0}^{2}k_{0}^{2}/4$ obtained with the pure harmonic $\eta\left(  x\right)
=\eta _{0}\cos k_{0}x$~\cite{deGennes03}. Comparison with experimental substrate profiles determined by Hitchcock \emph{et
al.}~\cite{roughExp} shows that (\ref{eq:eta}) can be used to
realistically represent actual rough substrates. For example, given the experimental profile
in \Figr{01}b (solid line of the upper plots), $k_{0}$ is readily determined from 
$k_0 = 2\pi n/\sqrt{5/3}$ and $\eta_0$ from $\eta_0 = \sqrt{\langle \eta^2 \rangle}$, 
where $n$ corresponds to the number of maxima per unit length~\cite{Rice52} and 
$\langle \cdot \rangle$ denotes an ensemble average over all
substrate realizations. To obtain the dashed profile as an approximation
to the experimental profile using \eqref{eq:eta}, the finite length of the profile is matched
to the period of \eqref{eq:eta} to get $N=26$, which then allows us to determine
the $\alpha_m$ and $\beta_m$ by projecting the experimental profile onto
their corresponding harmonics. On the other hand, the lower plot in \Figr{01}b 
is generated with random $\alpha_m$ and $\beta_m$ with $N=1000$, where $\eta_0$ and $k_0$
are kept the same so that the lower plot belongs to the same substrate family with the upper one. 
For the statistical analysis that follows, a large number of substrate realizations,
typically $20,000$, will be utilized.

Different substrate descriptions might have been used that have been observed in
profilometry measurements, as for example representations that
exhibit statistical self-affinity, whose spectral density follows
the power law $\propto k^{2D-5}$~\cite{Rough}, where $1<D<2$ is the
fractal dimension. The differentiability requirement of the
substrate representation together with the fact that in reality a
self-affine structure cannot persist for all lengthscales requires
imposing lower and upper wavenumber cutoffs. However, as we shall
see later the results for the statistics of the contact lime motion
are qualitatively the same regardless of the representation we
employ.  

\begin{figure}[ptb]
\includegraphics[scale=\zoom]{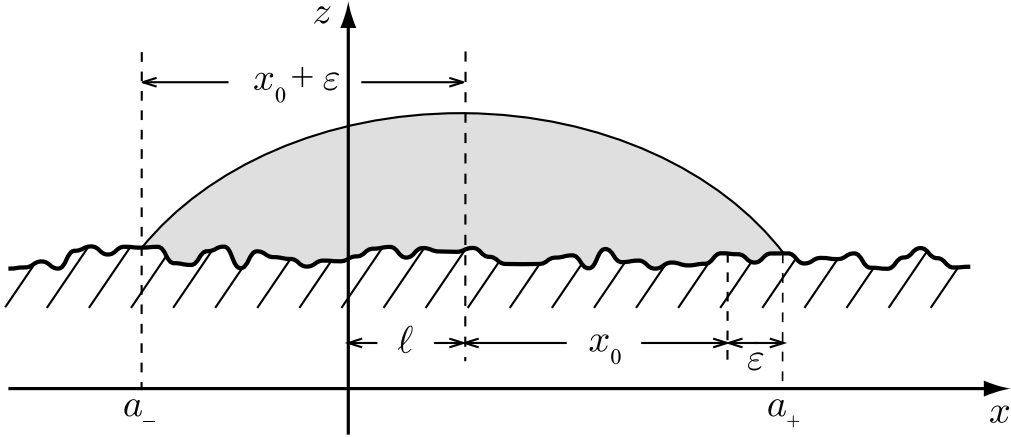}\caption{Droplet lying between $a_{-} \leq
x \leq a_{+}$ on a random topographical substrate. The droplet
shift, $\ell= \tfrac{1}{2} (a_{+}+a_{-})$, is the distance the
droplet midpoint moves away from $x=0$ and the contact line
fluctuation, $\varepsilon= \tfrac{1}{2} (a_{+}-a_{-})-x_{0}$,
measures deviations of the contact line location away from the
flat-substrate radius,
$x_{0}(t)$.}%
\label{fig:02}%
\end{figure}

To facilitate the analysis, we introduce the
\emph{contact line fluctuation} $\varepsilon$ and \emph{droplet shift} $\ell$
along the substrate (see Fig.~\ref{fig:02}), defined in terms of the contact
line locations as
\begin{equation}
\varepsilon=\tfrac{1}{2}\left(  a_{+}-a_{-}\right)  -x_{0}\quad\text{and}\quad%
\ell=\tfrac{1}{2}\left(  a_{+}+a_{-}\right)  ,\label{eq:dfn_eps}%
\end{equation}
where $x_{0}$ is the contact line location when the droplet spreads
on a flat substrate symmetrically about $x=0,$ which approaches
$\sqrt{3}$ in the long time limit. The droplet shift is a measure of
the distance the droplet midpoint is displaced from $x=0,$ whereas
the contact line fluctuation measures the deviation of the droplet
radius from the flat-substrate radius. As the amplitude of the
topographical features is taken to be small, we also expect
$\varepsilon \ll1$. By linearizing about the flat-substrate
equilibrium we obtain
\begin{gather}
 \varepsilon=\frac{3\eta_{0}}{2\sqrt{N}}\sum_{m=1}^{N}\left(  \alpha_{m}%
\sin\lambda_{m}\ell+\beta_{m}\cos\lambda_{m}\ell\right) {\cal
I}(\sqrt{3}\lambda
_{m}),\label{eq:eps} \\
 \sum_{m=1}^{N} \left(  \alpha_{m}\cos\lambda_{m}\ell-\beta_{m}\sin\lambda
_{m}\ell\right)  {\cal J}(\sqrt{3}\lambda_{m}) =0,\label{eq:ell}%
\end{gather}
\begin{figure}[ptb]
\includegraphics[scale=\zoom]{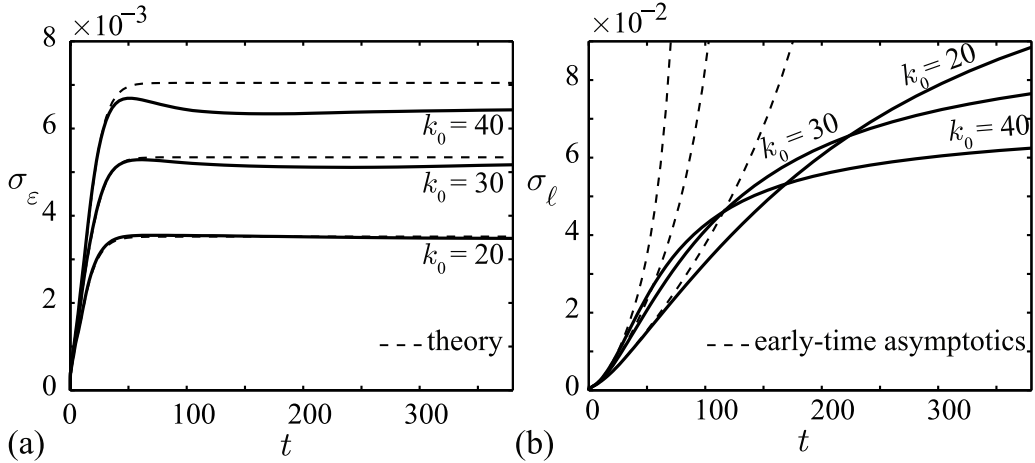}\caption{Dynamics of droplet spreading over
randomly varying substrates with $\eta_{0}=5\times10^{-4}$  and
$k_{0}=20$, 30 and 40 using 20,000 samples for each substrate
family. (a) Standard deviation of $\varepsilon$ as a function of
time. The numerical experiments (solid line) are indistinguishable
from theory (dashed line) for $k_{0}=20$. (b) Standard deviation of
$\ell$ as a function of time. The early-time asymptotics agree with
numerical experiments up to $t\sim \mathcal{O}(30)$. For large times
the solid lines asymptote at values
predicted from the long-time analysis.}%
\label{fig:03}%
\end{figure}where $\lambda_{m}=k_{0}m/N$, ${\cal I}(x) = \sinc x-\cos x -(x/3) \sin
x$ with $\sinc x=x^{-1} \sin x$ and ${\cal J}(x) = x\cos x-\sin x$.
By looking at the neglected terms in this linearization procedure,
we find that (\ref{eq:eps}) and (\ref{eq:ell}) accurately predict
droplet equilibria provided that $\eta_{0}k_{0}^{2}\ll1$. To conform
with this condition, together with our principal aim to examine the
effects of small scale roughness, we focus on substrate families
with $\eta_{0}\ll1$ and $1\ll k_{0}\ll\eta_{0}^{-1/2}$. Hence, for a
droplet with $L=0.5\mathrm{mm}$, $\alpha_s=15^{\circ}$ and substrate
topographies with amplitudes $0.5\mu\mathrm{m}$
$(\eta_0\approx4\times10^{-4})$, $\eta_0k_0^2<1$ for $l_0> 77\mu$m
(beyond the region of wavelengths in~\Figr{01}b, but Hitchcock
\emph{et al.} give materials with smaller amplitudes and larger
wavelengths, e.g. for silica, $\eta_0 \sim 0.006 - 3.1 \mu$m, $l_0
\sim 65 - 110 \mu$m). At these scales, slip is more important than
intermolecular forces in controlling the spreading dynamics, as the
study by Hocking on droplet spreading over flat substrates has
demonstrated ~\cite{Hocking94}. By the central limit theorem, we
also see from \eqref{eq:eps} that $\varepsilon$ is approximately a normally
distributed random variable. Contributions to its mean are of
$\mathcal{O} \left( \eta_{0}\right)$ and
originate from $\left\langle \alpha_{m}\sin\lambda_{m}\ell+\beta_{m}\cos\lambda_{m}%
\ell\right\rangle $. If all equilibria are taken into account
however, this quantity vanishes and contributions to the mean are of
$\mathcal{O}(\eta_0^2)$. Therefore, to fully assess the effects of
spatial heterogeneities on wetting, the equilibria attained from the
droplet dynamics need to be considered instead.

For arbitrary $t$, the IDEs obtained for deterministic substrates
in~\cite{SavvaKall09} are appropriately modified to model random
spatial heterogeneities described by~(\ref{eq:eta}) and are
linearized for $\varepsilon \ll 1$. This calculation is rather
involved and lengthy. The final equations are of the form,
\begin{eqnarray}
\dot{\varepsilon}+{\cal A}\left(  t\right)  \varepsilon & = & \frac{\eta_{0}}{\sqrt{N}}%
\sum_{m=1}^{N} \left(  \alpha_{m}\sin k_{m}\ell  +  \beta_{m}\cos k_{m}\ell\right)  {\cal B}\left(  t,k_{m}\right),  \nonumber \\
\dot{\ell} & = & \frac{\eta_{0}}{\sqrt{N}}\sum_{m=1}^{N}  \left(
\alpha_{m}\cos k_{m}\ell-\beta_{m}\sin k_{m}\ell\right) {\cal
C}\left(t,k_{m}\right), \nonumber
\end{eqnarray}
%\begin{align}
%&\dot{\varepsilon}\!+\!A\left(  t\right)  \varepsilon\!=\!\frac{\eta_{0}}{\sqrt{N}}%
%\!\sum_{m=1}^{N}\! \left(  \alpha_{m}\sin k_{m}\ell \! + \! \beta_{m}\cos k_{m}\ell\right) \! B\left(  t,k_{m}\right),  \nonumber \\
%&\dot{\ell}=\frac{\eta_{0}}{\sqrt{N}}\sum_{m=1}^{N}  \left(  \alpha_{m}\cos
%k_{m}\ell-\beta_{m}\sin k_{m}\ell\right)  C\left(t,k_{m}\right), \nonumber
%\end{align}
where ${\cal A}(t)$, ${\cal B}(t,k_m)$ and ${\cal C}(t,k_m)$ are
complicated functions of their arguments that we omit here for the
sake of brevity. Their time-dependence enters through $x_0$ and its
time derivative with $x_0$ satisfying, $3\dot{x_0}\ln [2x_0/(\lambda
e^2)] = 27x_0^{-6}-1$. The linearity of the equation for
$\varepsilon$ also implies that this quantity is a normal variable
for all times, whose variance may be computed explicitly. Figure
\ref{fig:03}a shows plots of the standard deviation of
$\varepsilon$, $\sigma_{\varepsilon}$, as a function of time for
$\eta_{0}=5\times10^{-4} $ and $k_{0}=20$, 30 and 40. When
$k_{0}=20,$ the theoretically predicted curve is indistinguishable
from the one obtained from numerical experiments, but as expected
the agreement tends to degrade as the condition
$\eta_{0}k_{0}^{2}\ll1$ no longer holds. Determining the
time-evolution of the standard deviation of $\ell$, $\sigma_{\ell}$,
explicitly is a formidable task due to the highly non-linear nature
of the equation for $\ell$, but the early-time behavior can be found
by linearizing about $\ell=0$.  In Fig.~\ref{fig:03}b we show the
evolution of $\sigma_{\ell}$ as computed from numerical experiments,
together with the early-time behavior predicted by the linear theory
for the same parameters as in Fig.~\ref{fig:05}b. There is excellent
agreement up to $t\sim \mathcal{O}\left(  30\right)  $; after that
time the theoretically predicted variance goes to infinity since the
linearized equation for $\ell$ predicts a Cauchy variable in the
long-time limit. Comparing the time-scales over which
$\sigma_{\varepsilon}$ and $\sigma_{\ell}$ saturate, reveals that
the droplet `footprint', $2\left(  \varepsilon+x_{0}\right)  $,
approaches equilibrium over a shorter timescale compared to the time
for $\ell$ to reach equilibrium, which suggests that the droplet
slides without spreading along the various substrate features to
find the final equilibrium position.
\begin{figure}[ptb]
\includegraphics[scale=\zoom]{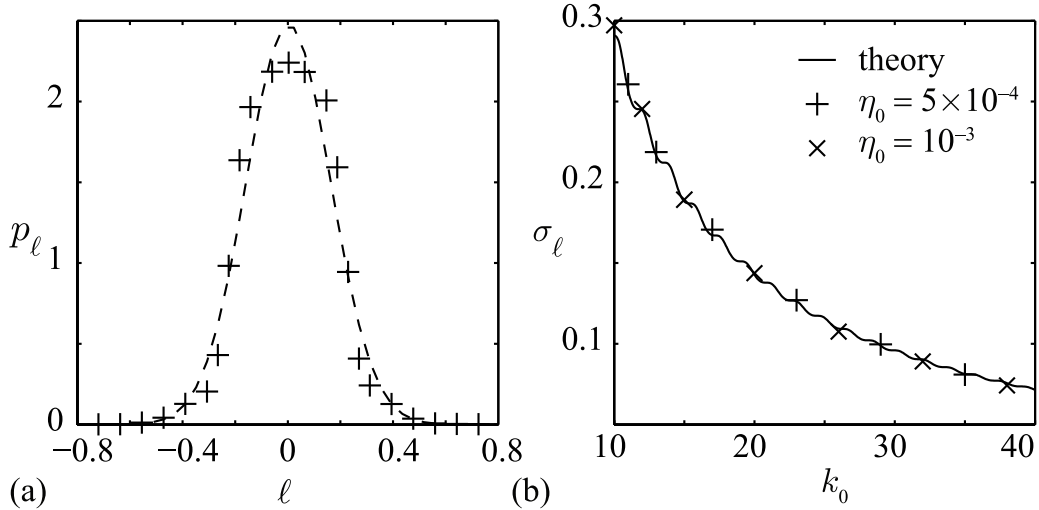}\caption{Statistics of $\ell$ for $k_{0}$
ranging from 10 to 40 with $\eta_{0}=5\times10^{-4}$,$(+)$ and $10^{-3}$,
$(\times)$, taken over 20,000 samples and $N=1,000$. (a) Probability density function of $\ell$
when $\eta_{0} = 5\times10^{-4}$ and $k_{0}=20$ compared with a normal density
of the same variance (dashed line); (b) Standard deviation of $\ell$ as a
function of $k_{0}$, illustrating the excellent agreement of (\ref{eq:var_ell}%
) with numerics.}%
\label{fig:04}%
\end{figure}

In the long-time limit, we cannot solve for $\ell$ explicitly since
\eqref{eq:ell} is nonlinear and moreover it admits infinitely many
solutions. However, the evolution towards equilibrium fixes the
solution to \eqref{eq:ell} to be the stable equilibrium that is
closest to $\ell=0$, a problem which is reminiscent of the highly
nontrivial `first-passage problem' in probability
theory~\cite{Rice52}. Interestingly, Fig.~\ref{fig:04}a reveals that
the probability density of $\ell$ as $t\rightarrow \infty$,
$p_{\ell}$, is far from being a normally distributed random variable
as the comparison with the density of the normal variable with the
same variance demonstrates. By taking into account the mean distance
between zeros of (\ref{eq:ell}) \cite{Rice52}, together with the
fact that on average half of the closest equilibria are unstable, we
find that
\begin{equation}
\sigma_{\ell}^2  =\tfrac{5}{6}\pi^{2}k_{0}^{-2}\left(
1-\tfrac{1}{2}\sinc 2\sqrt{3}k_{0}\right)  +\mathcal{O}\left(
k_{0}^{-4}\right) \label{eq:var_ell}%
\end{equation}
in the limit $k_{0}\gg1$. It is clear that as the substrate features become
rougher the droplet has a tendency to shift/slide less along the substrate.
This behavior is confirmed in Fig.~\ref{fig:04}b, where we plot the
theoretically predicted $\sigma_{\ell}$ as a function of $k_{0}$ together
with numerical experiments for different substrate families, confirming also
the independence of $\ell$ on $\eta_{0}$.

\begin{figure}[ptb]
\includegraphics[scale=\zoom]{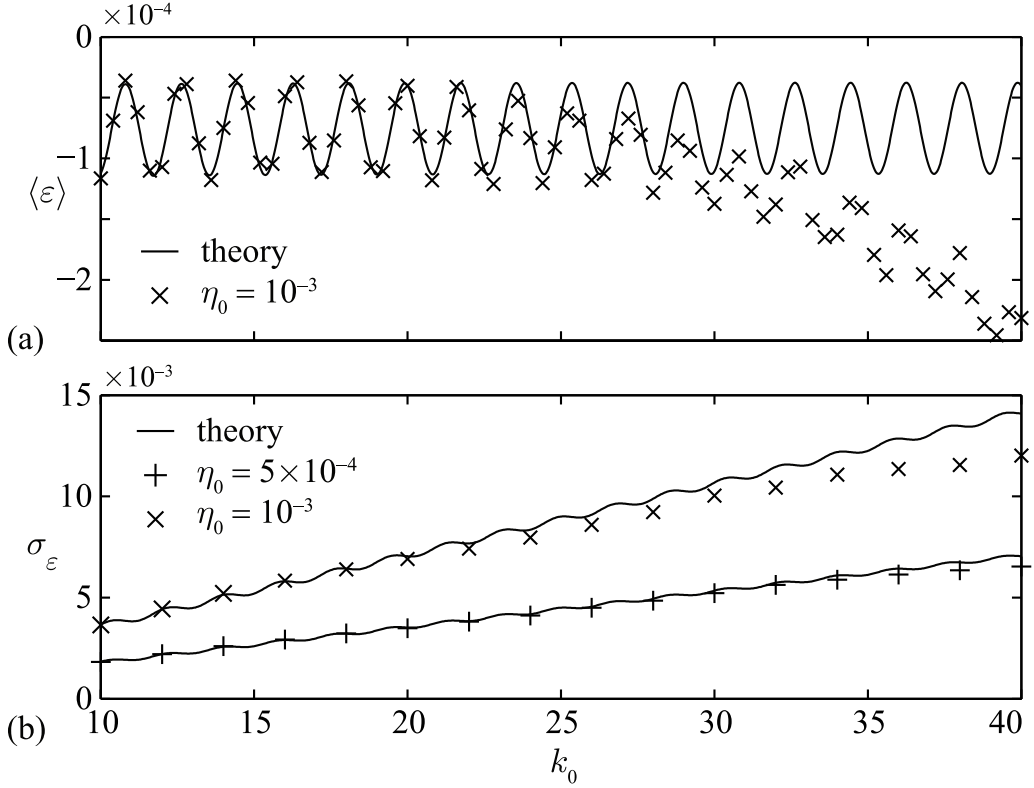}\caption{Statistics of $\varepsilon$ for the same
parameters as in Fig.~\ref{fig:04}. (a) Comparison of the
numerically determined $\left\langle \varepsilon\right\rangle $ and
$\left\langle \varepsilon\right\rangle _{\mathrm{approx}}$ as a function of
$k_{0}$; (b) Comparison of the theoretical and numerical standard deviation of
$\varepsilon$ as a function of $k_{0}$. The agreement is excellent for
$\eta_{0}k_{0}^{2}\ll1$.}%
\label{fig:05}%
\end{figure}

From our numerical experiments we also found that $\langle
\varepsilon \rangle <0$ in the long-time limit, thus suggesting that
surface roughness reduces wetting. Such behavior appears to
contradict Wenzel's theory but it signifies the fact that the
droplet has to overcome the energy barriers that separate the
multiple equilibrium droplet states. This effect is demonstrated in
the recent experiments of Chung \emph{et al }~\cite{Chung07}, where
spreading perpendicular to the grooves of parallel-grooved
substrates violates Wenzel's law, and further supported by the work
of Cox~\cite{Cox83} on wedge equilibria over (deterministic)
three-dimensional rough substrates, who postulated that
roughness-induced wetting enhancement is due to a higher-order
effect which manifests itself when spreading does not occur
perpendicular to the substrate grooves. A semi-analytical expression
for $\left\langle \varepsilon\right\rangle $ can be
obtained by noting that from our numerical experiments, $\left\langle \alpha_{m}\sin\lambda_{m}%
\ell+\beta_{m}\cos\lambda_{m}\ell\right\rangle =F\left( k_{0}\right)
{\cal J} (\sqrt{3}\lambda_{m})/(k_{0}\sqrt{3})$, where $F$ appears
to depend weakly on $\eta_{0}$ and $k_{0}$ and equals to 3 for
$\eta_{0} k_{0}^{2}\ll1$. Based on
this observation, $\left\langle \varepsilon\right\rangle $ is found to be%
\begin{equation}
\left\langle \varepsilon\right\rangle _{\mathrm{approx}}\approx-\tfrac
{3}{8}\eta_{0}\left(  2-\cos2\sqrt{3}k_{0}\right)  +\mathcal{O}\left(
\eta_{0}k_{0}^{-1}\right)  ,\label{eq:mean_eps}%
\end{equation}
suggesting that $\left\langle \varepsilon\right\rangle $ has an
oscillatory behavior as $k_{0}$ varies. Furthermore, the mean
apparent contact angle is now larger than the static contact angle
by an amount $2|\left\langle \varepsilon\right\rangle|/\sqrt{3}$ to
leading order in $\varepsilon$. Figure~\ref{fig:05}a depicts a plot
of (\ref{eq:mean_eps}) as a function of $k_{0}$ together with the
mean obtained in numerical simulations for substrates with
$\eta_{0}=10^{-3}$, considering 20,000 samples from each substrate
family. For smaller $k_{0},$ the agreement between the semi-analytic
approximation and the numerical experiments is evident, but as the
substrate becomes more rough so that $\eta_{0}k_{0}^{2}\ll1$ is no
longer valid and nonlinear effects become appreciable, there is a
clear deviation towards a progressive reduction of the mean droplet
radius. This implies that the apparent contact angle tends to
increase with substrate roughness, thus pointing towards a
substrate-induced, hysteresis-like effect.

The variance of $\varepsilon$ can be deduced from (\ref{eq:eps}) by converting
the Riemann sum into an integral by taking $N\rightarrow\infty$,
\begin{equation}
\sigma_{\varepsilon}^2  =\tfrac{1}{8} \eta_{0}^{2}k_{0}^{2}%
\left(  1-3\sinc 2\sqrt{3}k_{0}\right)  +\mathcal{O}%
\left(  \eta_{0}^{2}\right)  ,\label{eq:var_eps}%
\end{equation}
when $k_{0}\gg1$. The theoretically predicted $\sigma_{\varepsilon}$ is in very good
agreement with the simulated one
as shown in Fig.~\ref{fig:05}b, where we plot $\sigma_{\varepsilon
}$ as a function of $k_{0}$ when $\eta_{0}=5\times10^{-4}$ and
$\eta_{0}=10^{-3}$. Had we used a self-affine substrate
representation, the results are qualitatively the same since the
leading order variance of $\varepsilon$ differs from the leading
order of \eqref{eq:var_eps} only by a factor
$3(2-D)(g^{4-2D}-g^{2})/[(D-1)(1-g^{4-2D})]$, that depends on two
additional parameters, namely the fractal dimension $D$ and on the lower
to upper cutoff wavenumber ratio, $g$. Numerical studies of
$\sigma_{\ell}$ also confirm qualitative agreement.

To conclude, we have presented the first detailed and systematic
investigation of the motion of 2D droplet fronts over randomly
varying shallow substrates by using a model derived first
principles, i.e. from the governing hydrodynamic equations in the
limit of small contact angles. Analytical predictions made for the
equilibria of the two quantities of interest, the contact line
fluctuation $\varepsilon$ and droplet shift along the substrate
$\ell$ exhibit excellent agreement with numerical experiments.
However, our analysis revealed that considering droplet equilibria
alone cannot fully determine the effects of random substrates on
wetting; instead, their stability from the dynamical spreading
problem needs to be taken into account as well. For arbitrary times,
examination of the evolution of $\ell$ and $\varepsilon$ suggests
that on average the droplet has the tendency to slide without
spreading along the substrate before reaching equilibrium. In the
long-time limit, $\ell$ and $\varepsilon$ scale with $\sigma_
{\varepsilon}^2 \sim\eta_{0}^{2}k_{0}^{2}/8$ and $\sigma_{\ell}^2
\sim5\pi^{2}k_{0}^{-2}/  6 $, respectively. We believe that these
results will motivate further analytical and experimental studies on
the role of heterogeneities on wetting hydrodynamics.

We thank the anonymous referee for suggesting the statistical
self-affine substrate representation. We acknowledge financial
support from EPSRC Platform Grant No. EP/E046029.

\end{document}